\documentclass{scrartcl}
\usepackage{url}
\usepackage{graphicx}
\usepackage{amssymb}
\usepackage{color}

\begin{document}
\title{Operators for Space and Time in BeSpaceD}

\author{
Jan Olaf Blech and Keith Foster
}

\date{RMIT University, Melbourne, Australia}

\maketitle
\begin{abstract}
In this report, we present some spatio-temporal operators for our
BeSpaceD framework. We port operators known from functional
programming languages such as filtering, folding and normalization on abstract
data structures to the BeSpaceD specification language. We present the
general ideas behind the operators, highlight implementation details
and present some simple examples.
\end{abstract}

\section{Introduction}
This report continues our work on BeSpaceD by introducing new
operators for filtering, folding and normalization of BeSpaceD
descriptions. The new operators are inspired by operations from
functional programming languages and are applied to spatio-temporal
models. 

BeSpaceD is our framework for spatio-temporal reasoning \cite{bespaced1,bespaced0}.  BeSpaceD is
characterised by
\begin{itemize}
\item
A description language that is based on abstract datatypes.
\item 
Means to reason about descriptions formulated in the description
language.
\end{itemize}

Related work comprises process
algebra based 
formalisms \cite{cardelli03,cardelli04} and \cite{haar}. Similar to
this work, we have
created a
formal specification language as part of our BeSpaceD framework. In our case, specifications are
instances of abstract datatypes and follow a more functional
programming style. 
Further related is work on type systems in connection
with this process algebra work that has been introduced in
\cite{cairestypes}. Analog to the work presented in \cite{cairestypes}, BeSpaceD can also be used to
define spatio-temporal types. A verification tool to check
properties based on the process algebra inspired formalism is
described in \cite{slmc}. Applications
include concurrency
and ressource control. Further related are approaches for the
specification of hybrid systems. A framework for specifying
hybrid programs with stochastic features is presented in
\cite{platzer2011}.
Other logic approaches to spatial reasoning can be found, e.g., in \cite{hirschkoff,Bennett,zilio}. 
Specialised solutions for reasoning about geometric constraints
have been developed for robot path planning. This area has already been studied for
decades, see e.g., \cite{robots1,robots2}.  

This report provides an overview on BeSpaceD in
Section~\ref{sec:bes}. Filtering is covered in
Section~\ref{sec:filt}. Folding of time and space is covered in
Section~\ref{sec:fold} while normalization is described in
Section~\ref{sec:norm}. Section~\ref{sec:concl} concludes the report.

\section{BeSpaceD}
\label{sec:bes}
BeSpaceD is a spatio-temporal modeling and reasoning framework. Here,
we describe the modeling language, BeSpaceD-based reasoning, and
provide some implementation background information.

\subsection{BeSpaceD-based Modeling}
BeSpaceD is implemented in Scala. Its core functionality runs in a
Java environment. In the past, we have successfully applied BeSpaceD in different contexts such as  decision
support for factory automation~\cite{etfa,etfa2015}, coverage analysis for
mobile devices~\cite{han2015} and for verification of spatio-temporal
properties for industrial robots~\cite{apscc,fesca2014,cyberbeht}.

BeSpaceD models are created using Scala case classes. Thus, we provide a functional abstract datatype-like feeling.
Major language constructs which are currently available (see also \cite{2015arXiv151204656O}) are provided below.
An {\tt Invariant} is the basic logical entity, something that is
supposed to hold for a system throughout space and time. Invariants
can and typically do, however, contain conditional parts, something
that requires a precondition to hold, e.g., a certain time implies a
certain state of a system. 
Constructors for basic logical operations connect invariants to form a
new invariant. Some of these basic constructors are provided below:
{\small
\begin{verbatim}
abstract class Invariant;

abstract class ATOM extends Invariant;

case class OR (t1 : Invariant, t2 : Invariant) extends Invariant;
case class AND (t1 : Invariant, t2 : Invariant)  extends Invariant;
case class NOT (t : Invariant) extends Invariant;
case class IMPLIES (t1 : Invariant, t2 : Invariant)  extends Invariant;

case class BIGOR (t : List[Invariant]) extends Invariant;
case class BIGAND (t : List[Invariant]) extends Invariant;

case class TRUE() extends ATOM;
case class FALSE() extends ATOM;
\end{verbatim}
}
\noindent Additional predicates are used to indicate timepoints,
timeintervals, events, ownership, and related information. 
{\small
\begin{verbatim}
...
case class TimePoint [T] (timepoint : T) extends ATOM; 
case class TimeInterval [T](timepoint1 : T, timepoint2 : T) extends ATOM; 
...
case class Event[E] (event : E) extends ATOM;
case class Owner[O] (owner : O) extends ATOM;
case class Prob (probability : Double) extends ATOM; 
case class ComponentState[S] (state : S) extends ATOM;
\end{verbatim}
}
\noindent Some geometric constructs are provided below:
{\small
\begin{verbatim}
case class OccupyBox (x1 : Int,y1 : Int,x2 : Int,y2 : Int) extends ATOM;
case class OccupyBoxSI (x1 : SI,y1 : SI,x2 : SI,y2 : SI) extends ATOM; 
case class EROccupyBox 
   (x1 : (ERTP => Int),y1 : (ERTP => Int),
    x2 : (ERTP => Int),y2 : (ERTP => Int)) extends ATOM; 
case class OwnBox[C] (owningcomponent : C,x1 : Int,y1 : Int,x2 : Int,y2 : Int) 
            extends ATOM; 
...
case class Occupy3DBox (x1 : Int, y1: Int, z1 : Int, 
             x2 : Int, y2 : Int, z2 : Int) extends ATOM;
...
case class OccupyPoint (x:Int, y:Int) extends ATOM
case class OwnPoint[C] (owningcomponent : C,x:Int, y:Int) extends ATOM

case class Occupy3DPoint (x:Int, y:Int, z: Int) extends ATOM
...
case class OccupyCircle (x1 : Int, y1 : Int, radius : Int) extends ATOM; 
case class OwnCircle[C] (owningcomponent : C, 
             x1 : Int, y1 : Int, radius : Int) extends ATOM;
...
\end{verbatim}
}
\noindent In addition, we have topological constructs which are shown below:
{\small
\begin{verbatim}
case class OccupyNode[N] (node : N) extends ATOM
...
case class Edge[N] (source : N, target : N) extends ATOM 
case class Transition[N,E] (source : N, event : E, target : N) extends ATOM 

\end{verbatim}
}

In summary, the language constructs comprise basic logical operators (e.g., {\tt
  AND}, {\tt OR}) and constructs for space, time, and topology.
For instance, {\tt OccupyBox} refers to a rectangular two-dimensional geometric
space. It is parameterized by its left lower and its right upper
corner points.  
An example is provided below: if we want to express that the
rectangular space with the corner points $(42, 3056)$ and $(1531,
2605)$ is subject to a semantic condition ``A'' between
integer-defined time points $100$ and $150$, we may use the following BeSpaceD formula:
{\small
\begin{verbatim}
...
  IMPLIES(AND(TimeInterval(100,150),
               Owner("A")),
      OccupyBox(42,3056,1531,2605))
...
\end{verbatim}
}

\subsection{BeSpaceD-based Reasoning}
BeSpaceD comes with a variety of library-like functionality that
involves the handling of specifications. 
BeSpaceD formulas can be efficiently analyzed, i.e.,  verifying spatio-temporal and other properties.
A simple example involves the specification of a point in time and a
predicate. BeSpaceD can derive
the spatial implications from these definitions.
We have implemented algorithms and connected tools such as an external
SMT solvers (e.g., we have a connection to z3 \cite{z3}). These can help to resolve geometric constraints such as overlapping of different areas in time and space.
Different operators (for instance, breaking geometric constraints on
areas down to geometric constraints on points) exist. In this report, we are
looking at filtering in datatypes, normalization and spatio-temporal
fold operations. Filtering and normalization were already supported in
previous versions but have seen some updates in the
implementation. The folding functionality is new to this report.

\subsection{Implementation}
In the implementation of the presented operations we are using the Scala programming language. In
particular Scala’s case classes are used to define the language and we
make heavy use of pattern matching on types to implement
algorithms. For unit testing we use {\it ScalaTest}~\footnote{\url{http://www.scalatest.org/}} which allows us
to easily code tests with assertions and rapidly re-run them as
needed. 
The team used a combination of the Eclipse ``Scala
IDE''\footnote{\url{http://scala-ide.org/}} (mostly for runtime
execution) and IntelliJ IDEA~\footnote{\url{https://www.jetbrains.com/idea/}} (mostly for for editing and refactoring).

\section{Filtering}
\label{sec:filt}
Filtering allows the selection of relevant information out of larger
BeSpaceD invariants thereby returning a sub-invariant.

Filter functions have the following signiture scheme:
\begin{verbatim}
filter(inv : Invariant,filtercondition) : Invariant 
\end{verbatim}
An invariant is filtered following a filtercondition. The filter
condition may be a spatial area or a timeinterval. All relevant
information for the filtercondition is encapsulated in the returned
invariant.

A variety of different filter functions are possible, each one
evaluating the invariant with respect to different levels of semantic depth.

\subsection*{Implementation}
As part of the implementation of a fold, a filter function is
required. To illustrate one interesting example of this we will
discuss the filterTime function. Presented below is one of our
versions of filtering time which is of particular interest because it
uses an elegant ''rewrite then simplify'' algorithm that greatly
reduces the code complexity compared to other versions of
filterTime. 
However, we did not perform an analysis of algorithmic complexity so far.

The key line of source code is:
\begin{verbatim}
case tp: TimePoint[IntegerTime] => 
           if (withinTimeWindow(tp, startTime, stopTime)) tp 
           else successReturn.not
\end{verbatim}
This line rewrites the time point predicate as either a true or false value.
"tp" effects no change to the invariant and "successReturn.not" will rewrite it as either a TRUE or FALSE value.
This depends on the parent context of the re-write: FALSE for conjunctions and TRUE for disjunctions.
Later in the simplification process:
\begin{verbatim}
  simplifyInvariant(filteredInvariant)
\end{verbatim}
will cause whole branches of the abstract datatype tree to be truncated where
the time point is rewritten as false because one of the standard
simplification rules is:
\begin{verbatim}
  case IMPLIES(FALSE(),t2) => TRUE()
\end{verbatim}
In effect this filters out all the sub expressions that do not match
the desired time point. Other aspects of the source code can be found below:
{\small
\begin{verbatim}

  def filterTime2(startTime: IntegerTime, 
                  stopTime: IntegerTime, 
                  successReturn: BOOL)
                     (invariant: Invariant): Invariant =
  {
    val filter = filterTime2(startTime, stopTime, successReturn) _

    def withinTimeWindow(timePoint: TimePoint[IntegerTime], 
                         startTimeInclusive: IntegerTime, 
                         stopTimeExclusive: IntegerTime): Boolean =
    {
      val time = timePoint.timepoint

      time >= startTimeInclusive && time < stopTimeExclusive
    }

    val filteredInvariant: Invariant = invariant match
    {
      case tp: TimePoint[IntegerTime] => 
                 if (withinTimeWindow(tp, startTime, stopTime)) tp 
                 else successReturn.not

      case IMPLIES(premise, conclusion) => 
              IMPLIES(filter(premise), filter(conclusion))

      case AND(t1, t2) => AND(filter(t1), filter(t2))
      case BIGAND(sublist) => BIGAND(sublist map filter)

      case OR(t1, t2) => OR( filter(t1), filter(t2) )
      case BIGOR(sublist) => BIGOR(sublist map filter)
   
      case other => other
    }

    simplifyInvariant(filteredInvariant)
  }
\end{verbatim}
}

\section{Folding}
\label{sec:fold}
\begin{figure}
\centering
\includegraphics[scale=0.8]{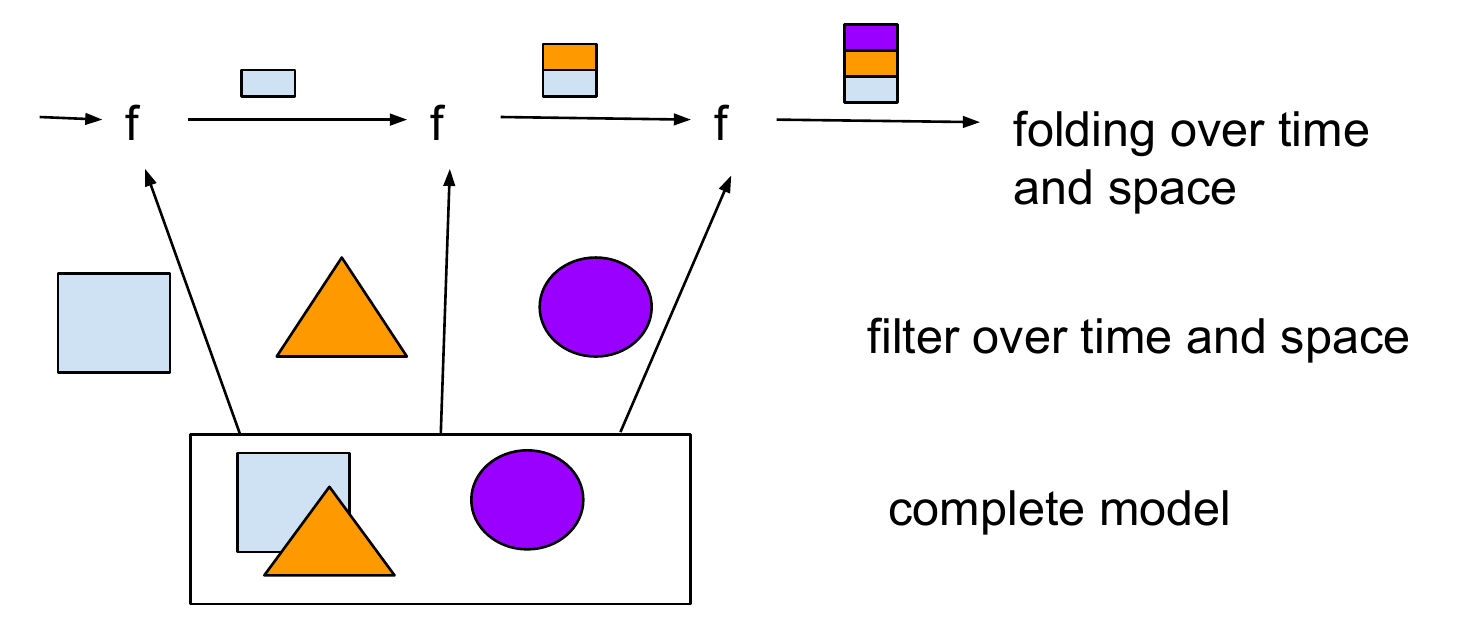}
\caption{Fold operator for space and time}
\label{fig:foldop}
\end{figure}

Figure~\ref{fig:foldop} shows an illustration of our fold operation
for time and space. The general principle of folding, the iteration
through time and space while accumulating processed information is shown.
In this section, we discuss the folding of time and space in separate
subsections and follow our implementation.

\subsection{Time}
A generalized signature for a folding time function is provided below. No assumptions are made on the implementation of time, other than that time has to be partially ordered.
{\small
\begin{verbatim}
foldTime[A,T] (invariant : Invariant, a: A,starttime : T,
                       stoptime : T,step: T, f [A -> Invariant -> A])
\end{verbatim}
}

\subsection*{Implementation}

We are exemplifying the folding of time by using an example with
weather data. We are regarding a
geometric space (a matrix-like structure) that contains values
indicating cloud coverage of an area.
To validate folding time we devised a complex test case that aggregates clouded areas over multiple time points.
This test case uses a naive approach as it requires a strict format for the knowledge invariants.
The knowledge about the space and whether its cloudy or not is encoded
as follows:
{\small
\begin{verbatim}
// Time
val t1 = new IntegerTime(1)
val t2 = new IntegerTime(2)
val t3 = new IntegerTime(3)

// Overlapping Boxes
val b1 = OccupyBox(1, 1, 10, 10) // Area 100
val b2 = OccupyBox(5, 5, 15, 15) // Area 121
val b3 = OccupyBox(10, 10, 20, 20) // Area 121

// Time occupied
val to1 = IMPLIES(TimePoint(t1), b1)
val to2 = IMPLIES(TimePoint(t2), b2)
val to3 = IMPLIES(TimePoint(t3), b3)

val timeSeries = List(to1, to2, to3)
val conjunction = BIGAND(timeSeries)
\end{verbatim}
}

In order to calculate the area occupied for one time step we
implemented a function called {\tt addAreaOccupied}. This function takes two parameters:
\begin{enumerate}
\item
total: This is called the accumulator as it will store the running total (an Integer) as the fold is being called recursively.
\item
item: This is BeSpaceD data (an {\tt invariant}) that represents the knowledge of the spatial cloud system.
\end{enumerate}
{\small
\begin{verbatim}

def calculateArea(box: OccupyBox): Int =
   Math.abs((box.x2 - box.x1 + 1) * (box.y2 - box.y1 + 1))

def addAreaOccupied(total: Int, item: Invariant): Int = {
  val area = item match {
    case IMPLIES(imp, box: OccupyBox) => calculateArea(box)
    case _ => 0
  }

  total + area
}
\end{verbatim}
}
It is interesting to note that the result of this function is a calculation derived from the two parameters (as seen in the last line above).
In this case the accumulator (total) is added to the area which is calculated from the second parameter (item).
This is the essence of an "aggregation function" that is commonly used
by all fold operations: The accumulated value and a function applied
to the next iteration
item are combined to give the next accumulated value.

The signature for an implemented function that folds time is as follows:
\begin{verbatim}
  def foldTime[A](
                   invariant: Invariant,
                   accumulator: A,
                   startTime: IntegerTime,
                   stopTime: IntegerTime,
                   step: IntegerTime.Step,
                   f: (A, Invariant) => A
                   ): A
\end{verbatim}

Additionally, for our example, we need to setup a few more parameters. The accumulator
used in the fold needs an initial value:
\begin{verbatim}
val initialValue = 0
\end{verbatim}
And finally we need to define a series of “steps” in time for the fold
operation to ``fold''. In this test case we specified the following:
\begin{verbatim}
t1, t3, 1
\end{verbatim}
This series of values simply represents start with time point 1, stop at time point 3 and step through time incrementing by 1.

Here is what the call to the foldTime function with the parameters
described above looks like:
\begin{verbatim}
val foldedTime = foldTime[Int](
  conjunction,
  initialValue,
  t1, t3, 1,
  addAreaOccupied
)
\end{verbatim}

To validate the test runs correctly we assert the expected result:
\begin{verbatim}
assertResult(expected = 342)(actual = foldedTime)
\end{verbatim}
The expected result of 342 is explained as follows:
There are three time points: 1, 2 and 3 each with their own boxes: b1, b2, b3.
The area of each are 100, 121, 121.
The fold of these three areas will be the sum which is 342.

\begin{figure}
\centering
\includegraphics[scale=0.5]{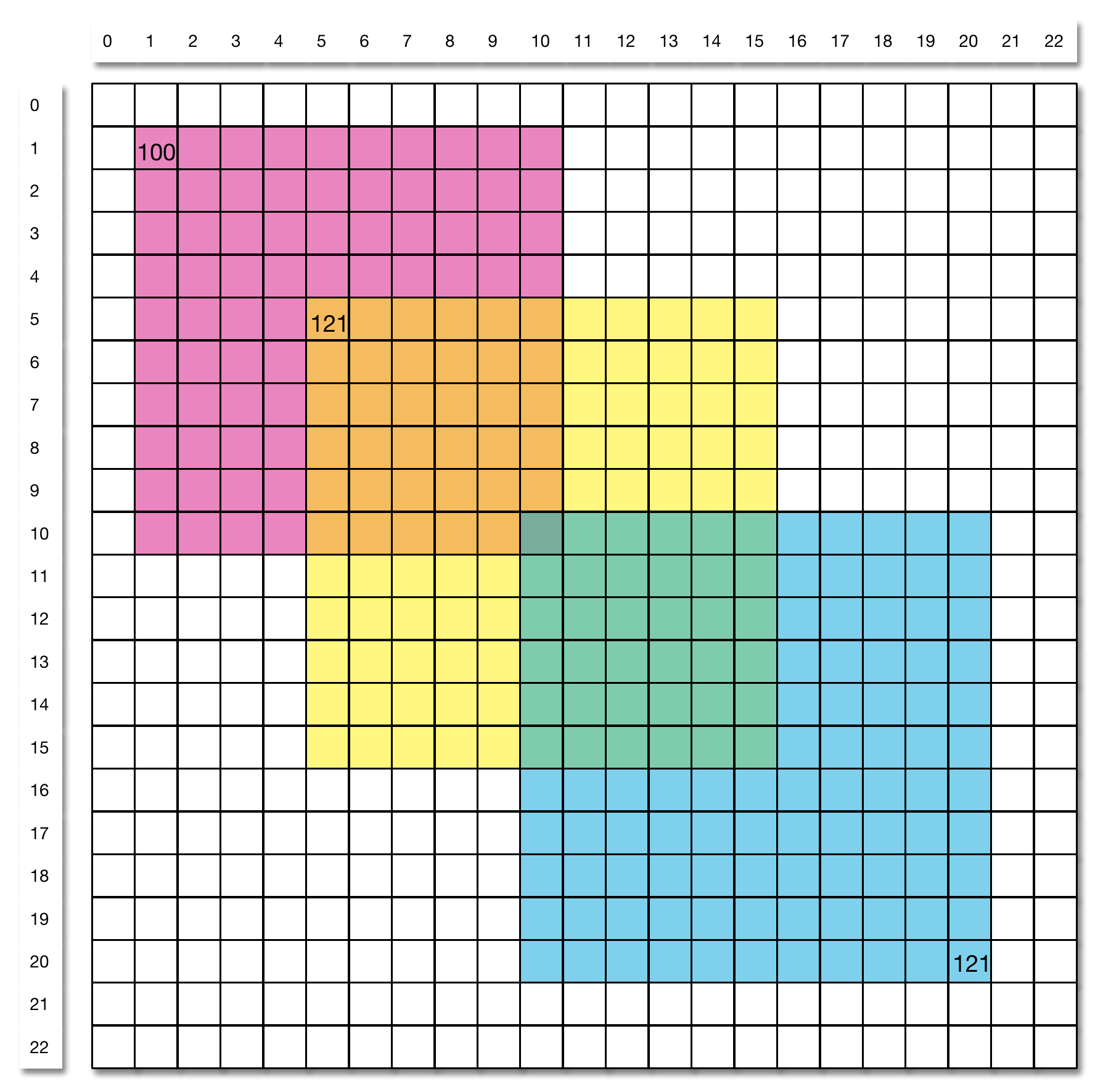}
\caption{Fold  over time illustrated}
\label{fig:foldtime}
\end{figure}

The actual fold over time example is also depicted in Figure~\ref{fig:foldtime}. The regarded spatial area stays the same, but the {\it cloud} moves over time (depicted in different colors).

\subsection{Space}

Folding space in general is done by using a function with the following signature. Note, that it works using generic area descriptions.

{\small
\begin{verbatim}
foldSpace[A] (
    invariant: Invariant,
    z: A, startarea: Invariant,
    stoparea: Invariant,
    stepareaOrBox: Invariant,
    f: (A -> Invariant -> A)) 
\end{verbatim}
}

\subsection*{Implementation}

A concrete signature for an implemented  function that folds space is as follows and assumes that areas are boxes:
{\small
\begin{verbatim}
foldSpace[A](
                  invariant: Invariant,
                  accumulator: A,
                  startArea: OccupyBox,
                  stopArea: OccupyBox,
                  translation: Translation,
                  f: (A, Invariant) => A
                  ): A
\end{verbatim}
}

To validate folding space we devised a simple test case that
aggregates clouded areas over multiple spatial iteration steps.
This test case uses a naive approach as it requires a strict format for the knowledge invariants.
The knowledge about the space and whether its cloudy or not is encoded
as follows:
{\small
\begin{verbatim}
    // Boxes
    val b1 = OccupyBox(1, 1, 10, 10)
    val b2 = OccupyBox(5, 5, 15, 15)
    val b3 = OccupyBox(10,10, 20,20)
    val b4 = OccupyBox(21,21, 30,30)

    // Space owner-occupied
    val s1 = IMPLIES(mountain, b1)
    val s2 = IMPLIES(cloud, b2)
    val s3 = IMPLIES(cloud, b3)
    val s4 = IMPLIES(mountain, b4)

    val spaceSeries = List(s1, s2 , s3, s4)
    val conjunction = BIGAND(spaceSeries)
\end{verbatim}
}

In order to apply the foldSpace function we need to pass in
the calculateArea function from the above example as the parameter f
and a spatial model as the invariant.

Additionally we need to setup a few more parameters. The accumulator
using in the fold needs an initial value:
{\small
\begin{verbatim}
val initialValue = 0
\end{verbatim}
}
And finally we need to define a series of steps�� in space for the
fold operation to ``fold''. In this test case we defined the following:
{\small
\begin{verbatim}
    val startBox = OccupyBox(1, 1, 5, 5)
    val stopBox  = OccupyBox(26,26,30,30)
    val step: Translation = (5, 5)
\end{verbatim}
}

This means we want the first box cover coordinates: (1,1) ... (1,5), (2,1) ... (2,5) ... (5,5).
In this example there are six iteration steps none of which overlap. In addition, as future work we would like to regard overlapping steps and explore semantic implications for this.

The code piece below shows a call to the foldSpace function with the parameters described above:
{\small
\begin{verbatim}
    val foldedSpace = foldSpace[Int](
      normalisedData,
      initialValue,
      startBox, stopBox, step,
      addCloudyArea
    )
\end{verbatim}
}
To validate the test runs we assert the expected result using the
following code line:
\begin{verbatim}
    assertResult(expected = 76)(actual = foldedSpace)
\end{verbatim}

The expected result on 76 is explained as follows:

The first iteration box (1,1 ... 5,5) is   indicated to contain a
mountain (see above). However, one point also contains a cloud. This point is counted for cloud coverage. We further iterate through the boxes.
The second iteration box (6,6 ... 10,10) is indicated to contain a
cloud. We continue iterating through the iteration path thereby
examining all 6 boxes in the path.
The total cloud covered area in the iteration path adds up to  1 + 25
+ 25 + 25 + 0 + 0= 76.

In order to calculate a ``degree of cloudiness'' for one spatial step
we implemented a function called {\tt addCloudyArea}. This function takes two parameters:
\begin{enumerate}
\item
total: This is called the accumulator as it will store the running total (an Integer) as the fold is being called recursively.
\item
invariant: This is BeSpaceD data that represents the model containing
clouds and other spatio-temporal information.
\end{enumerate}
{\small
\begin{verbatim}
def addCloudyArea(total: Int, invariant: Invariant): Int = 
{
  def isCloudyArea(owner: Owner[Any]): Boolean = owner == cloud

  def calculateArea(list: List[Invariant]): Int = {
    val areas: List[Int] = list map {
      inv: Invariant =>
        inv match {
          case IMPLIES(owner: Owner[Any], point: OccupyPoint) => 
             if (isCloudyArea(owner)) 1 else 0
          case IMPLIES(owner: Owner[Any], 
                       AND(p1: OccupyPoint, p2: OccupyPoint)) => 
             if (isCloudyArea(owner)) 2 else 0
          case IMPLIES(owner: Owner[Any], 
                       BIGAND(points: List[OccupyPoint])) => 
             if (isCloudyArea(owner)) points.length else 0
          case _ => 0
        }
    }
    areas.sum
  }

  val area = invariant match {
    case AND(t1, t2) => calculateArea(t1 :: t2 :: Nil)
    case BIGAND(sublist: List[Invariant]) => calculateArea(sublist)
    case _ => 0
  }

  total + area
}
\end{verbatim}
}
It is interesting to note that the result of this function is a calculation derived from the two parameters (as seen in the last line above).
In this case the accumulator (total) is added to the area which is calculated from the second parameter (invariant).

\begin{figure}
\centering
\includegraphics[scale=0.4]{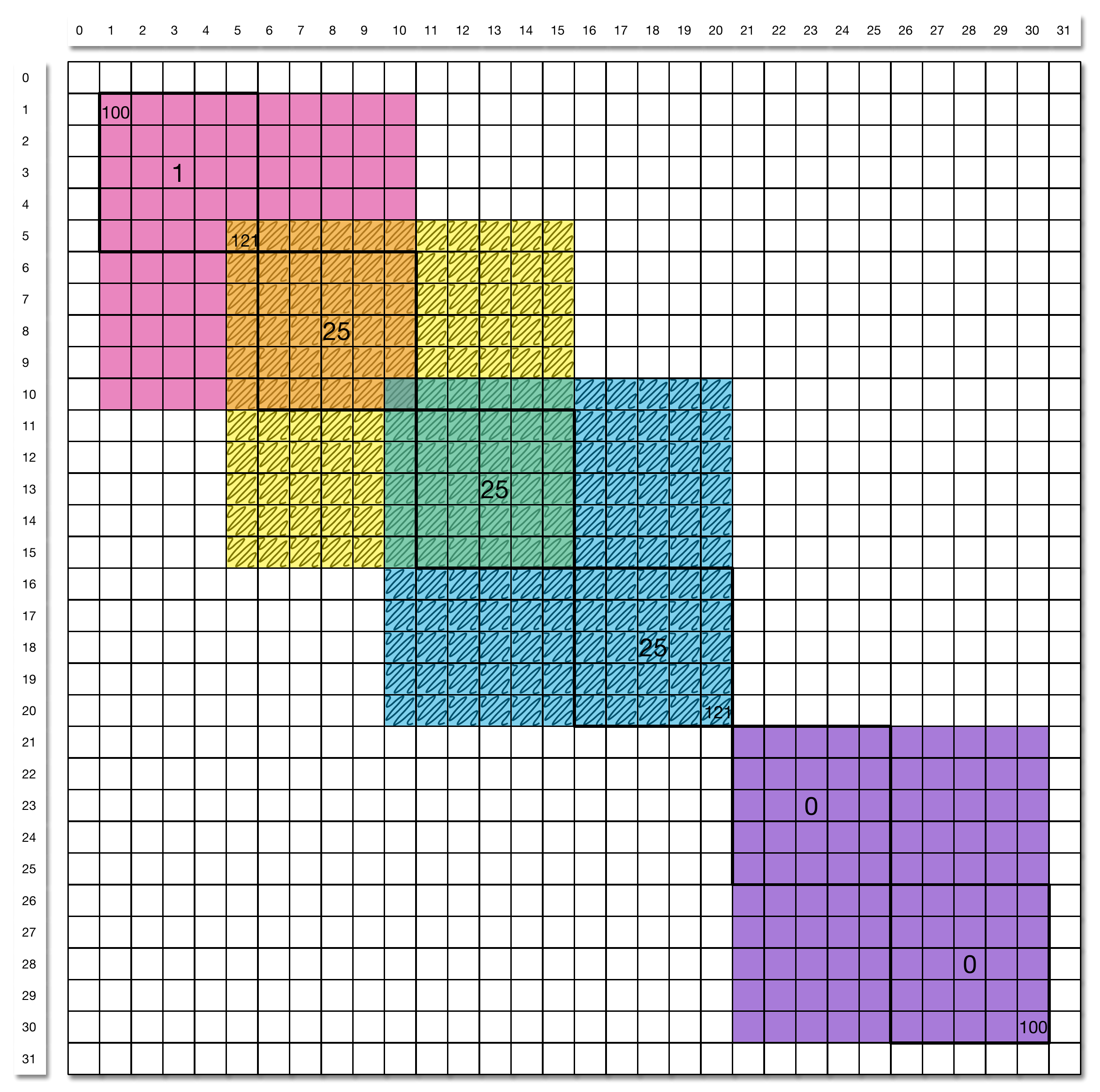} \\
\includegraphics[scale=0.4]{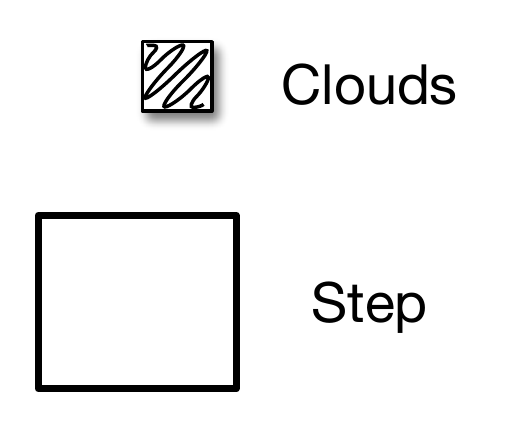}
\caption{Folding space example illustrated}
\label{fig:foldspaceex}
\end{figure}

The folding of space example is also illustrated in Figure~\ref{fig:foldspaceex}. Different boxes are shown for the different steps in our example.

\section{Normalization}
\label{sec:norm}
In order to make invariants comparable, we need to normalize
them. Normalization ensures, that the same invariants are
represented in the same way. However, there are different levels of
normalization. Normalization can take a higher or lower degree of
semantics into account. At the lower end, we may just reorder
arguments for logical operators such as ``and'' and ``or''. On the
other end, we may look into the semantic meaning of geometric shapes
and, e.g., replace areas with sets of points in order to make them
comparable. Different normalization strategies may require different
resources and work on different subsets of our language. Choosing an
appropriate form of normalization depends on the actual use-case. This is why we have not implemented a single normalization
function, but rather have a family of functions.

In general normalization involves term rewriting steps until a fix
point is reached. The invariant is then in a normal form. Several
properties of the term rewriting should be fulfilled such as confluence.
Comparison is then carried out by checking the resulting terms for equality.

\subsection*{Implementation}
To implement normalization in BeSpaceD we used functional composition
in chains. The type of each element in the chain is a function from
Invariant to Invariant:
\begin{verbatim}
type Processor[I, O] = I => O
type InvariantProcessor = Processor[Invariant, Invariant]
\end{verbatim}
We have created various invariant processors each of which perform one
step in the normalisation process. The following is a general
normalization function, showing the composition of four processors:
\begin{verbatim}
val normalize: InvariantProcessor = 
     flatten 
     compose 
     order 
     compose 
     deduplicate 
     compose 
     simplify
\end{verbatim}

A more specific example of a normalization function designed for data
of the form
\begin{verbatim}
Owner —>  OccupyPoints
\end{verbatim}
is described in the following.
It composes the standard normalization with a further processor
specific to the data structure.
\begin{verbatim}
val normalizeOwnerOccupied: InvariantProcessor = mergeOwners compose
normalize
\end{verbatim}
Below is a brief description of the step each processor does in the normalization process:
\begin{itemize}
\item
flatten -- rewrites nested conjunctions and disjunctions into single level conjunctions and disjunctions where possible.
\item
order -- orders the terms of all conjunctions and disjunctions according to a standard ordering.
\item
deduplicate -- removes duplicate terms of conjunctions and disjunctions.
\item
simplify -- uses term rewriting to rewrite expressions in a simpler
form. e.g. {\tt IMPLIES(TRUE, t2)} is rewritten as t2
\item
mergeOwners - expects a {\tt BIGAND} of {\tt IMPLIES} and conjuncts
all the conclusions of implications that have identical
premises. e.g. 
\begin{verbatim}
BIGAND(IMPLIES(A, X), IMPLIES(B, Y), IMPLIES(A, Z)) 
\end{verbatim}

is rewritten as 
\begin{verbatim}
BIGAND(IMPLIES(A, AND(X, Z)), IMPLIES(B, Y))
\end{verbatim}
\end{itemize}


\section{Conclusion}
\label{sec:concl}

We presented new operators for BeSpaceD in this paper. These operators
comprise filtering, folding and normalization for spatio-temporal
formulas. The operations are inspired by functional programming
languages and are ported to the spatio-temporal context. In addition
to theoretical considerations we presented an implementation and a
testing infrastructure for the BeSpaceD framework.

\subsection*{Acknowledgement}
The implementation work presented in this paper is open source and
freely available. It has
been done as part of the research and development activities in the Australia-India
Centre for Automation Software Engineering at RMIT University in Melbourne.

\bibliographystyle{eptcs}

\end{document}